**STELLAR OCCULTATIONS BY TRANSNEPTUNIAN OBJECTS: FROM PREDICTIONS TO OBSERVATIONS AND PROSPECTS FOR THE FUTURE.**

**J. L. Ortiz[1], B. Sicardy[2], J. I. B. Camargo[3,4], P. Santos-Sanz[1], F. Braga-Ribas[5,3,4]**
(1) Instituto de Astrofisica de Andalucia-CSIC, Glorieta de la Astronomia sn, 18008 Granada, Spain
(2) LESIA, Observatoire de Paris, Université PSL, CNRS, Sorbonne Université, Univ. Paris Diderot, Sorbonne Paris Cité, 5 place Jules Janssen, 92195 Meudon, France
(3) Observatório Nacional / MCTIC, Rua General José Cristino 77, 20921-400, Rio de Janeiro, Brazil
(4) LIneA, Rua General José Cristino 77, 20921-400, Rio de Janeiro, Brazil
(5) Federal University of Technology - Paraná (UTFPR/DAFIS), Av. Sete de Setembro, 3165, CEP 80230-901 - Curitiba - PR - Brazil

## ABSTRACT

**In terms of scientific output, the best way to study solar system bodies is sending spacecraft to make in-situ measurements or to observe at close distance. Probably, the second best means to learn about important physical properties of solar system objects is through stellar occultations. By combining occultation observations from several sites, size and shape can be derived with kilometric accuracy. Also, atmospheric properties can be derived if the body has an atmosphere. Furthermore, the technique can detect rings and even satellites (although rarely) around the main body. Except for the very special cases of Pluto and Charon, stellar occultations by Transneptunian Objects (TNOs) had never been observed until October 2009. This was because the ephemeris of the TNOs have much larger uncertainties than their angular diameters (typically of the order of ~10 milliarcsecond) and also because stellar catalogs were not accurate to the milliarcsecond level. Despite the difficulties, at the time of this writing, 43 occultations by 22 different Trans-Neptunian Objects, and 17 occultations by 5 Centaurs have been detected thanks to the efforts of several teams. Due to the complications of accurately predicting and observing these events, most of the successes have been achieved through wide international collaboration, which is a key issue to succeed in observing stellar occultations by TNOs. Multichord occultations are typically detected at a rate of ~3 per year on average, whereas the majority of the observed occultations are single-chord detections, which means that only one site detects the occultation. In these cases, no tight constraints on size and shape can be derived from those observations alone. Here we review most of the aspects involved in the complex process to successfully observe occultations, and present some of the lessons learned. There are good prospects for the future if we take advantage of the stellar catalog from the Gaia Data Release 2 and if we take advantage of dedicated observational programs of TNOs, to improve their orbits to the required level of accuracy, which is a critical aspect. All this, in combination with large and optimized telescope networks may allow an increase in the success rate and scientific output of the occultation technique applied to TNOs, in the short-term future.**

## 1. Introduction

The study of stellar occultations by Transneptunian Objects (TNOs) is a relatively new field because the first detection of a stellar occultation by a TNO was made only 9 years ago (Elliot et al. 2010). Although young, this field of work is quickly evolving and it is an essential one if we want to adequately characterize and understand the physical properties of the TNO population as a whole. This is because stellar occultations can provide extremely relevant information such as very accurate sizes and shapes, better than any other technique can do (except for spacecraft visits, which are extremely expensive, complex and require a lot of time). Size and shape are basic physical parameters that must be accurately known if we want to fully characterize a body, and are the first step toward deriving accurate densities. We need to know densities if we aim to determine internal compositions and make first guesses on the internal structure of solar system bodies (Carry et al. 2012).

On the other hand, once size and shape are correctly determined for a solar system body, combining this information with accurate brightness measurements allows deriving accurate geometric albedos, which are also an important piece of information to properly interpret many of the visible light observations of the TNOs, particularly their spectra (in order to derive surface composition, as surface reflectance composition models are highly degenerate unless the albedo is known e.g. chapters 6 and 20 in this book). And geometric albedo is also fundamental to analyze the temperatures and thermal behavior of the TNOs (see e.g. chapter 7). Even though measurements of the thermal output of solar system bodies can be used to obtain sizes and albedos (combining the thermal fluxes with visible light measurements), these so-called radiometric techniques have considerable limitations. For TNOs, sizes and albedos determined from radiometric techniques are only accurate to the 10 to 20% level (Mueller et al. 2010), and information on shape is not directly derived from the models. On the other hand, thermal measurements are extremely difficult to obtain for the TNOs because their peak of thermal emission is in the submillimeter range, around 100 μm, where our atmosphere blocks the electromagnetic radiation. Therefore, good radiometric studies in the past required the use of sophisticated space observatories with very limited life-time such as Spitzer and Herschel. Because of that, only a sample of around 130 TNOs have radiometrically determined sizes and albedos (Mueller et al. 2009). And unfortunately the accuracy of the resulting products is not high. Stellar occultations do not suffer those problems and by comparing the occultation-based with the radiometric-based results for a given body, the radiometric models or thermophysical models can be tuned to derive other thermal parameters that play a role in the thermal measured flux, such as thermal inertia, but also information on spin axis orientation, spin rate and other parameters can be retrieved or at least constrained using the combination of thermal data with occultation results (Mueller et al. 2018).

Besides, shapes derived from the occultations can be used to determine densities under the assumption of hydrostatic equilibrium of a homogeneous body, provided that the spin rate is known and some constraint on the spin axis orientation is used (Sicardy et al. 2011, Ortiz et al. 2012, Braga-Ribas et al. 2013, Benedetti-Rossi 2016, Schindler et al. 2017, Dias-Oliveira et al. 2017). Nevertheless, at least for the case of Haumea this approach fails to derive the actual density (Ortiz et al. 2017). This implies the interesting consequence that the object is not in hydrostatic equilibrium and/or is not homogeneous. Apart from all the above, stellar occultations provide a powerful means to determine the presence of atmospheres through the gradual immersion and emersion of the stars (during the disappearance and reappearance of the star, respectively) or through the study of central flashes. The presence or absence of atmospheres down to a small fraction of the microbar level of surface pressure is easily determined nowadays and this is an important topic because highly volatile ices can potentially generate partial or even full atmospheres in TNOs (Stern and Trafton 2008 and chapter 9 in this book). So far, no clear evidence for global atmospheres on a TNO other than Pluto have been found, although a local atmosphere might be envisaged (Ortiz et al. 2012 and Chapter 9).

Another key aspect of stellar occultations is the possibility to detect narrow and dense dust structures such as rings around the occulting bodies through the dimmings that they can cause in the stellar flux. The occultation detection of rings around the centaur Chariklo (Braga-Ribas et al. 2014) as well as the detection of a dust structure in Chiron with many similarities to Chariklo's rings (see Ortiz et al. 2015, although interpretation in terms dust shells is possible Ruprecht et al. 2015) represented an important breakthrough in planetary science, and the subsequent discovery of a ring around Haumea, one of the dwarf planets in the Transneptunian Belt, has also been a major discovery with plenty of implications (Ortiz et al. 2017), which opens a new field and raises many questions, including how frequent these structures are in the current TNO population (see Chapter 12). All this calls for more observations of stellar occultations.

As a summary, it is clear that the interest in stellar occultations by TNOs is extremely high and the topic deserves particular attention in a review paper. In this chapter we deal with general aspects of occultations by known TNOs and also include the Centaurs, which are not TNOs strictly speaking, but since they are closely related to them, it makes sense to address their occultations here too.

Serendipitous occultations by small unknown TNOs also provide a very powerful means of studying the Transneptunian region, because the technique has the potential to derive the number density of bodies of a given size range and their size distribution (Roques et al. 2006, Bickerton et al. 2008, Bianco et al. 2009, Wang et al. 2010, Schlichting et al. 2012, Chang et al. 2013, Zhang et al. 2013, Liu et al. 2015; Pass et al. 2018). However, this topic is not covered here.

## 2. General results from stellar occultations thus far and lessons learned

### 2.1. Difficulties of predicting and observing occultations by TNOs

Although the power of stellar occultations is well recognized in the planetary science community and has been widely exploited to characterize the asteroid belt, it was clear that succeeding in this endeavor would be challenging for the Transneptunian region, in contrast to stellar occultations by asteroids[1] (which are now relatively easy to predict and observe). Stellar occultations by TNOs have intrinsic problems: In order to predict if a TNO will occult a star, and given that the typical angular diameter of a 300-km TNO at 40 AU is 10 milliarcseconds (mas), one must know the positions of the stars in catalogs to that level of accuracy at least, and the ephemeris of the TNO must also be accurate to that level. Otherwise, predictions are extremely uncertain and thus useless. A TNO of that size at 40AU has a brightness of $m_V$=22 for an average geometric albedo of 0.08. For the stars, the positional accuracy is not a problem anymore thanks to the sub-mas accurate astrometry from the Gaia space mission (Prusti et al. 2016, Lindegren et al. 2018).

Unfortunately, typical positional uncertainties for the TNOs from orbits in the JPL Horizons database are in the order of 300 mas, and that is true for other sources of orbits such as the Minor Planet Center, Astorb or Astdys (see Fig. 1). The number of astrometric observations of a given distant small body, as well as the arc length (years) over which these observations are distributed, influences the accuracy of its ephemeris (Fig. 1).

[1] In this chapter we do not include the TNOs within the "asteroid" category, as asteroids have rocky compositions and are confined in the region inside 5.2 AU, whereas the TNOs are the progenitors of the short-period comets so their physical nature and characteristics are entirely different, and they reside much further away from the Sun. The IAU 2006 General assembly coined the term "small solar system body"' to refer to all the bodies that are not planets or dwarf planets. Some professional and mostly amateur astronomers use the term asteroid to collectively refer to all the bodies that are not planets, or comets (which display comas), and in that context TNOs could be asteroids, but we stick to the IAU terminology.

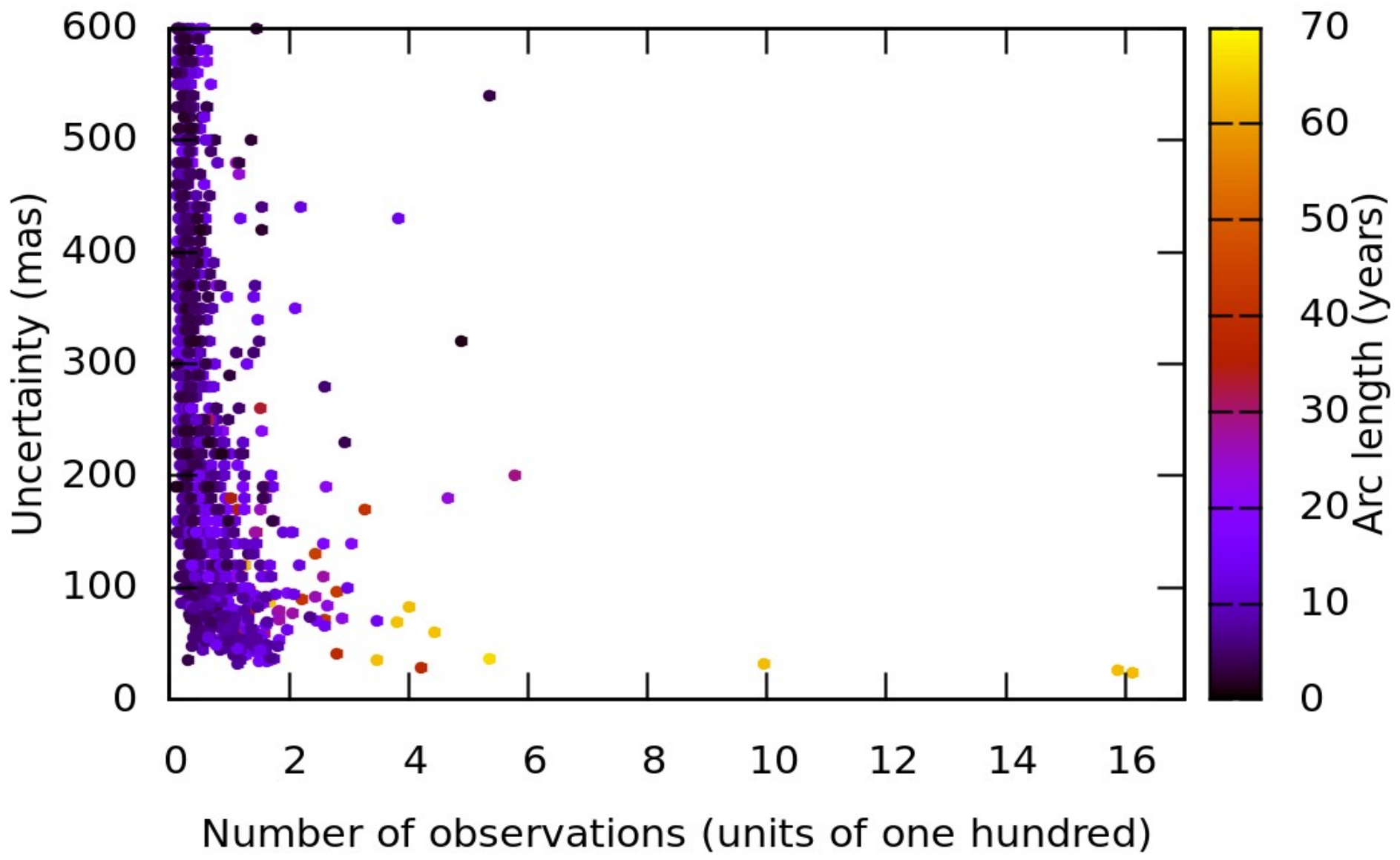

**Fig. 1** Uncertainties (Current 1σ Ephemeris Uncertainty – CEU) in ephemerides for objects with semimajor axes greater than 15 AU as a function of the number of observations and of the arc length over which these observations are distributed. Source of data: astorb.dat[2], as obtained on 01/OCT/2018. Only objects with an arc length greater than one year and more than 5 observations were considered. Uncertainties are calculated according to the Muinonen and Bowell (1993) formalism.

Our own statistics of monitoring 40 of the largest TNOs gives an rms deviation of the measurements with respect to the JPL orbits of 511 mas in right ascension and 273 mas in declination. Hence, we believe that 300 mas is a representative estimate of the uncertainties in the ephemeris, although obviously some objects have poorer orbits than others as shown in Fig. 1.

Given these overall numbers, it would appear nearly impossible to predict and observe occultations with a good success rate and not by mere chance after 30 or more "blind" attempts. However, there are ways to tackle this problem. The key issue is that even though the uncertainties in the absolute positions of the TNOs are as high as mentioned above, relative positions between the star to be occulted and the TNO can be obtained to the required accuracy of ~10 mas through specific observations. And once this is done, good enough predictions can be made. The approaches to make this possible are dealt with in section 3.

However, once an accurate prediction is made, we are still left with the problem of observing it. There are many difficulties at the observing stage. The region of the Earth where the occultation shadow path falls, weather, the brightness of the star and its elevation above the horizon at the time of the occultation, the lunar phase, organization logistics and contacts with the observers, training of the local observers (which are often newcomers in the occultation field), and many other complications play important roles. It is obvious that shadow paths on remote areas of the world, or mostly on oceans, result in observations not being practical. Also, if the occultations take place in areas of the Earth that are in sunlight, they will not be observable either. These aspects are probably the most evident difficulties, but there are lots of logistic aspects that can ruin observations. Some specific aspects on equipment are dealt with in sections 2.3, 4.1, 5.5 and 5.6.

## 2.2 General results

The evolution of the number of occultations that we are aware of (our own and those reported in amateur asteroid occultation networks, as well as in the scientific literature and conferences) observed since 2009, is shown in Fig. 2. Note that we did not include Pluto-Charon. Table 1 lists the occultations in our database. A version of this table that is continuously being updated can be found

[2] As obtained from ftp.lowell.edu, see also ftp://cdsarc.u-strasbg.fr/pub/cats/B/astorb/astorb.html.

in http://occresults.ga/results/

Table 1. List of observed stellar occultations by TNOs and Centaurs since the first TNO occultation in 2009 (not counting Pluto's system). The date of the events is in the first column. The second column indicates the name or the provisional designation of the TNO or Centaur. The third column indicates whether the event had or not involvement of the Paris Granada and Rio teams' collaboration (currently in the frame of the ERC Lucky Star project). The fourth column indicates if there is a scientific publication for the event.

| DATE | OBJECT | Lucky Star-RELATED | REFERENCE |
|---|---|---|---|
| 9 Oct 2009 | 2002 TX300 | no | Elliot et al. (2010) |
| 19 Feb 2010 | Varuna | yes | Sicardy et al. (2010) |
| 6 Nov 2010 | Eris | yes | Sicardy et al. (2011) |
| 8 Jan 2011 | 2003 AZ84 | yes | Dias-Oliveira et al. (2017) |
| 11 Feb 2011 | Quaoar | yes | Person et al. (2011) |
| 23 Apr 2011 | Makemake | yes | Ortiz et al. (2012) |
| 04 May 2011 | Quaoar | yes | Braga-Ribas et al. (2013) |
| 29 Nov 2011 | Chiron | no | Ruprecht et al. (2015) |
| 3 Feb 2012 | 2003 AZ84 | yes | Dias-Oliveira et al. (2017) |
| 17 Feb 2012 | Quaoar | yes | Braga-Ribas et al. (2013) |
| 26 Apr 2012 | 2002 KX14 | yes | Alvarez-Candal et al. (2014) |
| 25 Jun 2012 | Echeclus | no | |
| 15 Oct 2012 | Quaoar | yes | Braga-Ribas et al. (2013) |
| 13 Nov 2012 | 2005 TV189 | no | |
| 08 Jan 2013 | Varuna | yes | |
| 13 Jan 2013 | Sedna | yes | |
| 3 Jun 2013 | Chariklo | yes | Braga-Ribas et al. (2014) |
| 9 Jul 2013 | Quaoar | yes | |
| 29 Aug 2013 | Eris | yes | |
| 24 Nov 2013 | Asbolus | yes | |
| 2 Dec 2013 | 2003 AZ84 | yes | Dias-Oliveira et al. (2017) |
| 12 Dec 2013 | 2003 VS2 | yes | |
| 11 Feb 2014 | Varuna | yes | |
| 16 Feb 2014 | Chariklo | yes | Berard et al. (2017) |
| 1 Mar 2014 | Orcus/Vanth | yes | |
| 4 Mar 2014 | 2003 VS2 | yes | |
| 16 Mar 2014 | Chariklo | yes | |
| 29 Apr 2014 | Chariklo | yes | Leiva et al. (2017) |
| 24 Jun 2014 | Ixion | yes | |
| 28 Jun 2014 | Chariklo | yes | Leiva et al. (2017) |
| 07 Nov 2014 | 2003 VS2 | yes | |
| 15 Nov 2014 | 2007 UK126 | yes | Benedetti-Rossi et al. (2016) |
| 15 Nov 2014 | 2003 AZ84 | yes | Dias-Oliveira et al. (2017) |
| 26 Apr 2015 | Chariklo | yes | |
| 12 May 2015 | Chariklo | yes | Berard et al. (2017) |
| 3 Dec 2015 | 2002 VE95 | yes | |
| 12 Jun 2016 | Chariklo | yes | |
| 25 Jul 2016 | Chariklo | yes | Berard et al. (2017) |
| 08 Aug 2016 | Chariklo | yes | Berard et al. (2017) |
| 10 Aug 2016 | Chariklo | yes | Berard et al. (2017) |
| 10 Aug 2016 | Chariklo | yes | Berard et al. (2017) |
| 15 Aug 2016 | Chariklo | yes | Berard et al. (2017) |
| 20 Aug 2016 | Chariko | yes | |
| 1 Oct 2016 | Chariko | yes | Leiva et al. (2017) |
| 21 Jan 2017 | Haumea | yes | Ortiz et al. (2017) |
| 08 Feb 2017 | Chariklo | yes | |
| 7 Mar 2017 | Orcus/Vanth | no | Sickafoose et al. (2018) |
| 9 Apr 2017 | Chariklo | yes | |
| 20 May 2017 | 2002 GZ32 | yes | |
| 24 May 2017 | 2003 FF128 | no | |
| 22 Jun 2017 | Chariklo | yes | |
| 10 Jul 2017 | 2014 MU69 | no | Zangari et al. (2017) |
| 17 Jul 2017 | 2014 MU69 | no | Zangari et al. (2017) |
| 23 Jul 2017 | Chariklo | yes | |
| 24 Aug 2017 | Chariklo | yes | |
| 17 Nov 2017 | 2004 NT33 | yes | |
| 29 Dec 2017 | Bienor | yes | |
| 28 Jan 2018 | 2002 TC302 | yes | |
| 2 Apr 2018 | Bienor | yes | |
| 15 Jul 2018 | 2010 EK139 | yes | |
| 26 Jul 2018 | Quaoar | yes | |
| 4 Aug 2018 | 2014MU69 | no | |
| 2 Sep 2018 | Quaoar | yes | |
| 10 Sep 2018 | Varda | yes | |
| 19 Sep 2018 | 2002 KX14 | yes | |
| 28 Sep 2018 | 2004 PF115 | yes | |
| 20 Oct 2018 | 2015 TG387 | no | |
| 28 Nov 2018 | Chiron | yes | |
| 24 Dec 2018 | 2005 RM43 | yes | |
| 30 Dec 2018 | 2002 WC19 | yes | |
| 11 Jan 2019 | Bienor | yes | |
| 4 Feb 2019 | 2005 RM43 | yes | |

It should be pointed out that the majority of the occultations detected thus far have been single chord (observed from just one site). This situation does not allow us to determine sizes, but some information can still be derived (section 5). The number of multichord events is also shown in Fig. 2.

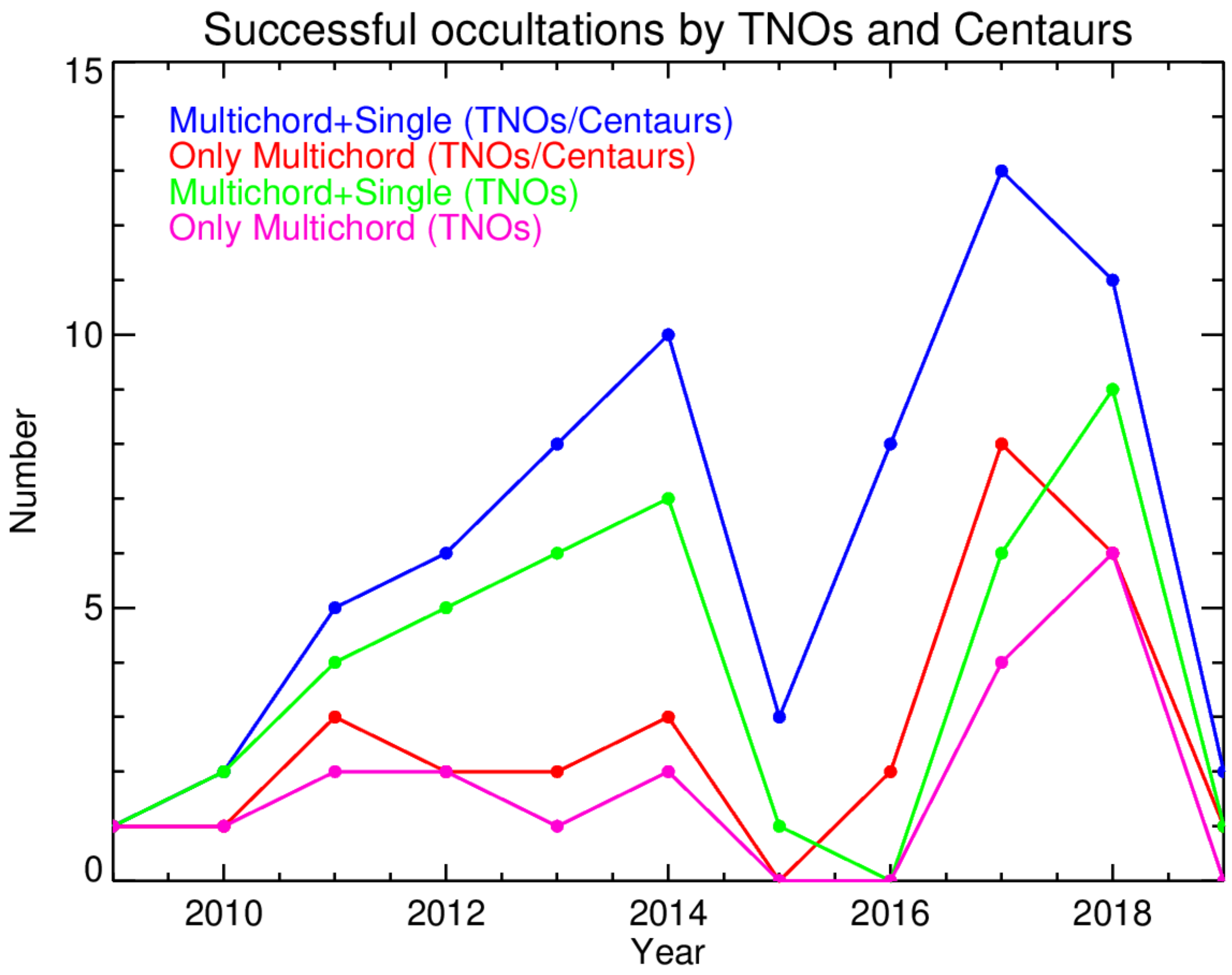

**Fig. 2** Number of detected occultations per year (TNOs plus centaurs) since 2009, Pluto excluded. In blue, all the events involving both TNOs and Centaurs. In green, only TNO events. In red, all the multichord occultations. In purple, multichord occultations involving only TNOs. Note that the data for 2019 contain only the first 2 months of the year, corresponding to the time of the revision of this paper. Hence the small number of detections in 2019.

Since 2009, there has been a general increase in the number of occultation events observed per year except for a fluctuation in 2015 and 2016. Note that 2019 has not finished at the time of the revision of this paper, so it can be expected that the final number of occultations in 2019 will increase. The number of multichord occultations per year has been steadier, with a noticeable increase in 2017 and 2018, possibly as a result of better astrometric updates thanks to the Gaia Data Release 2 (DR2), (see section 3).

There has been a wide variety of scientific results emerging from the occultations. For example, it is worth mentioning that a wide range of geometric albedos have been obtained. Also, a variety of shapes and sizes have been determined. Many of the results are still unpublished, partly because most of the events were single-chord. In table 2 some results from the already published events are summarized together with revisions of some values as explained in section 5.

Table 2. List of some physical parameters from the published stellar occultations and revised values with newer information on absolute magnitudes as described in the text. The Effective Diameter shown here is the diameter of a sphere with equal projected area. The values given are at the time of the occultation as published in the references.

| Object | Effective Diameter (km) | Geometric Albedo $p_v$ | Reference | Revised Effective Diameter (km) | Revised Geometric Albedo $p_v$ |
|---|---|---|---|---|---|
| 2002TX300 | 248±10 | 0.88±0.05 | 1 | **320±50** | **0.65±0.15** |
| Eris | 2326±12 | 0.96±0.03 | 2 | 2326±12 | **0.94±0.03** |
| Makemake | 1465±47 | 0.77±0.03 | 3 | 1465±47 | 0.77±0.03 |
| Quaoar | 1110±5 | 0.109±0.007 | 4 | 1110±5 | **0.112±0.01** |
| 2003AZ84* | 764±6 | 0.097±0.009 | 5 | 764±6 | **0.094±0.02** |
| 2007UK126 | 638±28 | 0.156±0.013 | 6,7 | **642±28** | **0.142±0.015** |
| Chariklo** | 268±12 | assumed | 8 | **268±12** | **0.038±0.008** |
| Haumea*** | 1392±26 | 0.51±0.02 | 9 | 1392±26 | 0.51±0.02 |

*The values given here are for the 2014 occultation

**The values given here are for the oblate spheroid model in the 2014-2016 time frame

***The effective diameter given is at the minimum of the rotational phase. The mean effective diameter (in equivalent area) is estimated at 1614 km.

1 (Elliot et al. 2010), 2 (Sicardy et al. 2011), 3 (Ortiz et al. 2012), 4 (Braga-Ribas et al. 2013), 5 (Dias-Oliveira et al. 2017), 6 (Benedetti-Rossi 2016), 7 (Schindler et al. 2017), 8 (Leiva et al. 2017), 9 (Ortiz et al. 2017).

For 2002TX300 the latest $H_V$ magnitude from Alvarez-Candal et al. (2016) has been used.

For Eris the latest $H_V$ magnitude from Alvarez-Candal et al. (2016) has been used. The effect of the satellite Dysnomia is negligible.

For Makemake, the effect of the discovered satellite is negligible.

For Quaoar the latest $H_V$ magnitude from Alvarez-Candal et al. (2016) has been used.

For 2003AZ84 the latest $H_V$ magnitude from Alvarez-Candal et al. (2016) has been used. Also the effect of the satellite has been taken into account by adding 0.01 mag in H.

For 2007UK126 the effect of the satellite has been taken into account by adding 0.032 mag in H.

For Haumea, no newer information is available.

For Chariklo we have used the minimum absolute magnitude of 7.35 in the historical time series when no ring contribution was present due to edge-on configuration (Duffard et al. 2014).

### 2.3 Some lessons learned

An important aspect is covering the shadow path with a sufficient number of observing stations. The larger the coverage with observers in the perpendicular direction to the shadow path, the higher the chances to succeed. Shadow paths on Earth are more frequently in the East-West direction than in the North-South direction because the apparent sky motion of the TNOs is mainly in this direction, but sometimes, near quadrature, where the TNO can move more in Declination than in Right Ascension we get shadow paths that move mainly North-South on Earth.

Note that the sky motion of the TNOs seen from Earth is basically due to parallax motion as the Earth travels in its orbit, because the orbital motions of the TNOs are very slow. As a result, typical speeds seen from Earth are in the range of 15 to 20 km/s (1 to 3 arcsec/hour in angular speed). For TNOs with sizes in the 200 km to 2000 km range, this means that occultations can last typically from 10 s to 100 s. If one can achieve a time accuracy of a tenth of a second in determining the disappearance and reappearance times, typical accuracies in the chord lengths are in the order of 1 to 3 km. Slow occultations near quadrature allow an even higher accuracy in the measured chords lengths, at least in theory. In practice, the final uncertainty in the timing of disappearance and reappearance of the star depends not only on clock time accuracy, but also (and primarily, in most of the cases) on the uncertainty of the photometry, which is the main parameter that plays a role in the square well models that must be fit to the occultation profiles (and from which the disappearance and appearance times are retrieved). This controls the final errors in the sizes of the chords and on the derived size for the TNO.

It is obvious that the fainter the stars, the more numerous they are, so the numbers of predicted occultations of faint stars are higher than those of brighter stars. For example, for a given TNO, there are 7 times more occultations of stars at magnitude 18 or brighter than at magnitude 15 or brighter (using the Gaia DR2 statistics of sources of 15 and 17.8 mag). Therefore, most of the occultation predictions will be for faint stars. However, detecting occultations of ~18-mag stars with widely available telescopes of diameter below ~40cm becomes difficult. A typical limit for the brightness of a star to observe occultations with a large enough number of telescopes is around 17.5 to 18. Fainter than that, only “large” telescopes with a diameter around 60cm or larger can contribute. Since the abundance of those telescopes on the surface of the Earth is much lower (or a fast deployment of telescopes in this size range is difficult), magnitude ~18 is currently a practical limit to pursue occultation observations in general, although in the future this might change (see the section on prospects for the future).

Even with a good prediction, on average, the success rate in our occultation program is typically 1 every 5.5 attempts. By success we mean a detection of the occultation from at least one site. Weather is the main reason ruining the campaigns, but also technical problems and observing mistakes are often present. Moreover, the number of participating stations is critical. Large campaigns involving at least 15 to 20 observing sites are typically needed to achieve multichord observations. From our own experience and from all the multichord stellar occultations published thus far, the campaigns that resulted in multichord detections typically involved >15 participating stations. The minimum to achieve a multichord detection was 15 stations, for Quaoar (Braga-Ribas et al. 2013). Hence, a number around 15 is probably the magical number of observing sites that need to be involved to guarantee success against weather, technical problems as well as other logistic issues and also give enough margin north and south (or east and west) of the nominal shadow path to accommodate changes in the shadow path within the typical uncertainties of the predictions. If predictions can be improved further than what we typically achieve, the number of participating stations may be decreased, but probably, a minimum number around 10 will still be needed. On average, attempts that do not reach this critical number of observing stations either fail or result in only single-chord or two-chord detections.

Most of the successes thus far have involved networks of telescopes from professional sites or specific networks established for TNO occultations, with some important contributions from amateur astronomers with access to high performance equipment, as most of the TNO events are not reachable with the broadly used setups by the amateur community.

From table 1, we can also see that most of the successful occultations were achieved thanks to the international collaboration of the groups at Paris, Granada and Rio. For the last few years this is under the umbrella of the Lucky Star project, a European Research Council (ERC) advanced grant of the H2020 program. Hence, international collaboration and coordinated efforts are of paramount importance at that point.

## 3. The future of the predictions

### 3.1. Techniques to make accurate predictions

As already mentioned, for many years prior to the existence of the Gaia DR2 catalog, the positions of the stars in the best astrometric catalogs were known at no better than ~50 mas accuracy for the most favorable cases (Assafin et al. 2012). To make matters worse, the average uncertainty in TNO ephemeris is of the order of 300 mas. Hence, it appears that accurately predicting stellar occultations would be impossible.

The approach thus far has been to focus on those TNOs that have the best orbits, which also usually happen to be the largest ones so their angular diameters are also the largest, and try to improve the orbits with accurate astrometry. But since the star catalogs to which the astrometric measurements had to be referred also had large errors in the past, in practice it was difficult to improve the orbits substantially. However, this strategy, in connection with case-to-case analysis and appropriate weighting of reported measurements to the Minor Planet Center and our own measurements can allow us to generate our own orbits, such as in the Numerical Integration of the Motion of an Asteroid (NIMA) scheme, see Desmars et al. (2015).

For the TNOs with the best orbits and the largest sizes, the chances of success are the highest so we often focused our predictions on them. Most of the results shown in Table 1 and Fig. 2 come from astrometric efforts concentrated on around 50 TNOs/Centaurs. Once potentially good events are identified (with the best possible orbits), the predictions must still be refined. This requires observing the TNO close in time to the occultation date from weeks to days prior to the occultation and getting astrometry with respect to the same reference frame as the star to be occulted, either with relative astrometry or using a high accuracy intermediate star catalog. Hence, good astrometry of the TNO with respect to the occulted star allows us to decrease the large positional uncertainties and we often achieve the typical 10 mas level needed.

An example of predicted shadow path after an astrometric update and the real shadow path on Earth is shown in Fig. 3. The uncertainty in telescopic CCD astrometric observations has two main components, one due to accuracy in determining the TNO centroid ($\sigma_c$) and another one due to the astrometric solution of the plates using reference stars in the field of view, $\sigma_s$. The two uncertainties add quadratically so that $\sigma_t^2=\sigma_c^2+\sigma_s^2$.

From our empirical tests on main belt asteroids with accurate orbits, the uncertainty of the TNO centroid determination is $\sigma_c$ ~1/2 FWHM/SNR where FWHM is the full width at half maximum of the stellar images and SNR is the signal to noise ratio achieved for the TNO (in fact, it can be shown that $\sigma_c$ = FWHM/(2.355*SNR) for well sampled, signal dominated sources (see e.g. Mighell 2005 for a detailed discussion). For $m_V$~22 TNOs under typical seeing (FWHM=1 arcsec), and a SNR=100, the uncertainty is 0.005” (5 mas). But apart from this error, as mentioned before, one has to add quadratically the error of the astrometric solution for the images $\sigma_s$, which typically goes as the square root of the number of stars

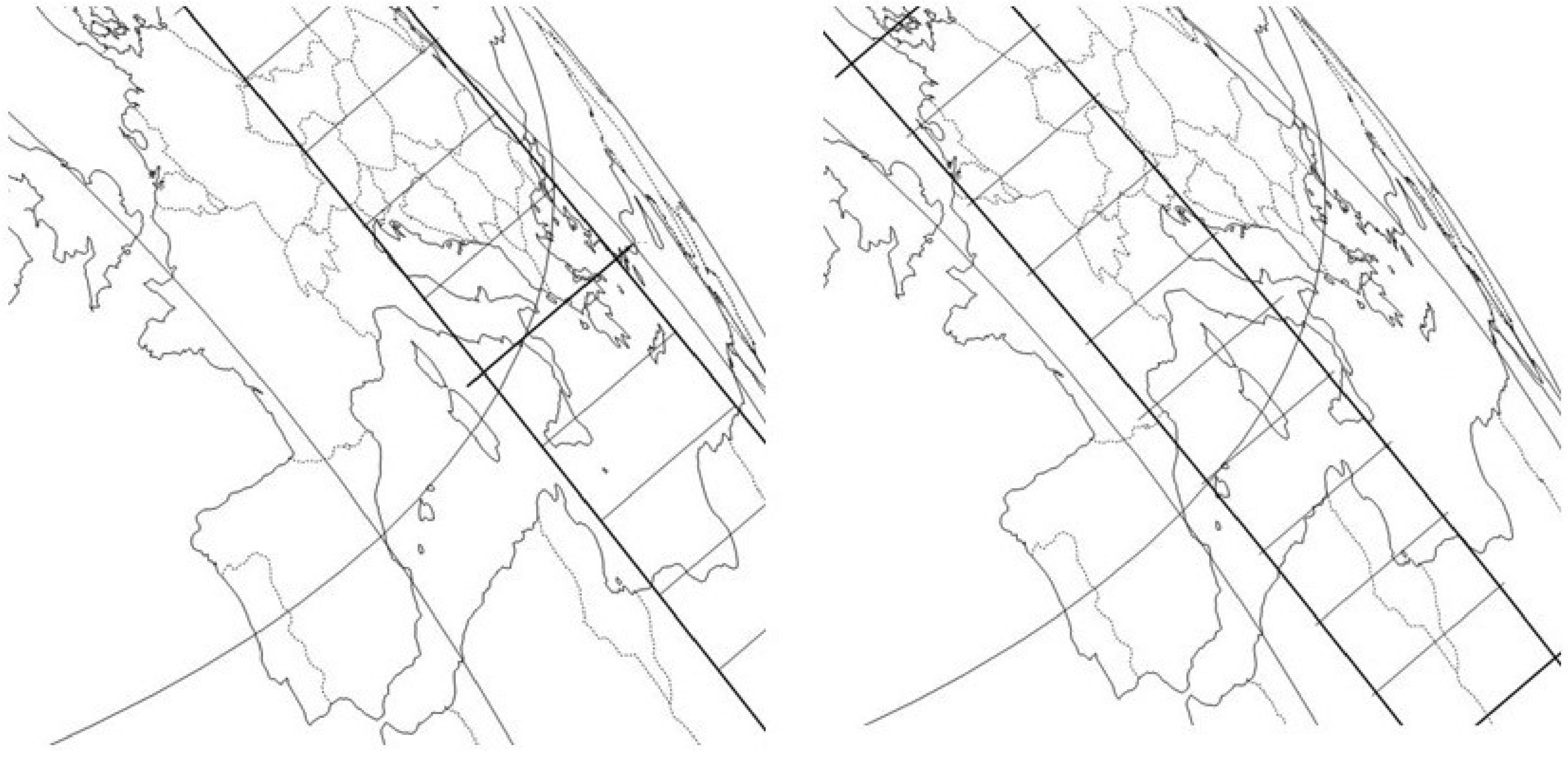

Fig. 3. Left: Prediction map of the stellar occultation by 2002TC302 on Jan 28, 2018, from an astrometric update several days prior to the event using images acquired with the 1.5m telescope at Sierra Nevada Observatory. The map shows the different countries as well as two meridians and a parallel. The straight lines delineate the star show path and the perpendicular lines to the shadow path indicate minutes. Right: The post-occultation reconstructed shadow path from a preliminary fit to the chords. Comparing the two figures and given that the shadow path has a width of 17 mas, it turns out that the prediction was good to the 10 mas level, with the prediction somewhat to the east compared with the reality. In this plot the shadow moves from south to north. Usually, most of the shadow paths of occultations by TNOs

are in the east-west direction.

used in the solution multiplied by the mean uncertainty in their positions. Thanks to Gaia DR2, we now have large enough number of stars even for small fields of view of the 2m-diameter or larger telescopes that can be used to derive astrometry. Therefore, for images with a sufficient number of stars, the required uncertainty of 0.01” can be obtained, provided that telescopes in the ~2m size range are used. Note that in order to obtain SNR=100 in typical ~1000s exposures, telescopes in the 2m size range are needed for TNOs with $m_V$~22.

Differential Chromatic Refraction in the atmosphere can also give rise to systematic errors. A strategy to minimize this is to make observations near transit and use filters, ideally red filters, or even better, near IR filters. As given in Stone (2002), these errors are below the few mas level at zenith angles of ~20$^o$, but depend on the color of the target and filter. Nevertheless, specific computations can be done to correct for this effect (Stone 2002). Another solution involves the knowledge of spectra for the TNO and the reference stars. There are other sources of errors for the TNO centroid, such as contamination from faint background stars, which can shift the centroid. Observations on several days are recommended in order to minimize the problem, which may be very important in crowded fields and at deep magnitudes. In these cases, PSF fitting is usually superior to regular centroid determination. Also, techniques of deblending and/or image differencing are possible solutions.

Binary TNOs or TNOs with satellites also require especial treatment because the centroid will not lie in general on the center of the primary but in between the primary and satellite, and will give rise to a systematic error. This has an important effect on Haumea, which had to be carefully addressed (Ortiz et al. 2017) and was also seen in Orcus (Ortiz et al. 2011). This is also a well-known effect on Pluto due to its satellite Charon and can be modelled efficiently (Benedetti-Rossi et al. 2014). In many cases, we do not know whether the TNO could have an unknown satellite large enough to bias the centroid determination. Another complication is the possibility that the occultation star could be a close binary so that the catalog coordinates do not really correspond to any of the stars. Sometimes this can be anticipated by analyzing the colors or through spectroscopic observations of the star. The star duplicity aspects will be covered in future Gaia data releases but not nowadays.

Since the advent of Gaia, the astrometric updates are much more easily done and have good accuracy provided that the aspects above are taken into account. This is probably increasing our success rate of the multichord events. This can be seen in Fig 2.

Before Gaia, the use of own star catalogues specifically tailored for the purpose of improving predictions of stellar occultations of several main TNO targets (Assafin et al. 2010, Assafin et al. 2012, Camargo et al. 2014, Young et al. 2008, Porter at al. 2018), or small “subcatalogs” around preselected occultation events have usually been the main approach to make accurate updates. For specific cases, this may still be necessary nowadays (for instance, to make predictions of occultations by very faint stars not in the Gaia catalog, to be observed with very large telescopes only).

TNOs that move in dense stellar fields also cause more occultations. We can take advantage of that, but if the stellar fields are too crowded and the TNO is faint, good prediction refinement is difficult because good astrometry cannot be easily carried out from the ground due to the contamination by background stars, unless long campaigns are devoted to the TNO, with plenty of telescope time.

### 3.2. The role of big telescopic surveys

As already stated in previous sections, an accurate prediction of stellar occultations by small distant Solar System Bodies relies on accurate stellar positions and ephemerides. Although the accuracy of stellar positions is no longer a problem, thanks to the sub-mas accurate astrometry from the Gaia space mission, distant small bodies still need improvement of their orbits, and this can be achieved to some degree by frequent observations with powerful enough telescopes.

One important point to be considered here is how fast the uncertainties in the ephemerides of these bodies increase, as a function of time, after the most recent observation used in the determination of their orbits. As an example (Fig. 4) we can use the Centaur (55576) Amycus. After a time span of 7 years without observations, the uncertainty in the position grows considerably.

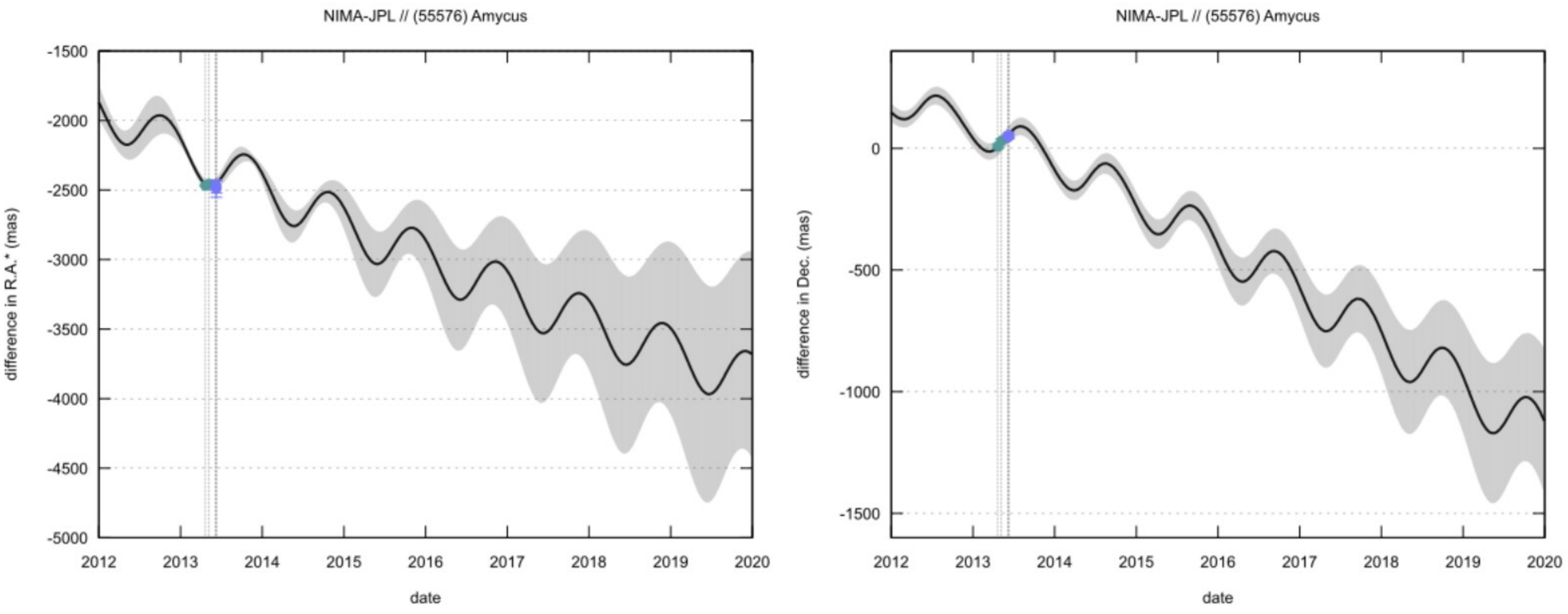

**Fig. 4** Differences (black wavy lines) in right ascension (left panel) and declination (right panel) between the positions of (55576) Amycus as obtained from the NIMA and JPL (version 7) ephemerides. The sense of the differences is NIMA minus JPL. The grey area is the uncertainty of NIMA at 1σ level. The green dot represents a Gaia DR2 based position of Amycus. The blue dot represents a Gaia DR1 based position of Amycus. Positions of this Centaur, from 1987, were taken from the Minor Planet Center and also used in the orbit determination.

The NIMA scheme mentioned earlier is a numerical integration procedure that has been successfully used to compute accurate orbits in order to predict stellar occultations (Desmars et al. 2015). It uses a specific weighting scheme for the input astrometry data that considers the individual precision of the observation, the number of observations obtained during one night by the same observatory, and the possible presence of systematic errors in positions. It is easily seen from Fig. 4 that the ephemeris uncertainty ranges from few tens of mas in those dates close to the most recent observations (around 2013.4) to hundreds of mas a few years later. Although the increase rate of the uncertainty depends on the observational history of each object and its orbit; it is clear that regular observations of the bodies are important to achieve and keep accurate orbits, provided that the input astrometry is accurate. The big large-area astronomical surveys, like the Dark Energy Survey (Flaugher 2005) and mainly the Large Synoptic Survey Telescope (LSST Science Book, Version 2.0, Abell et al. 2009), will play important roles in this regard (by providing their astrometric measurements of small solar system bodies to the Minor Planet Center).

As an example, the LSST, in particular, will record the entire sky visible from Cerro Pachón (Chile) twice a week and expects to observe around 40.000 TNOs over its ten years of operations, with more than one hundred observations for many of them. Single epoch (that is, non coadded) images should detect objects as faint as $m_r$ ~23.5. Positions as accurate as 10 mas should be obtained under suitable seeing and for the expected signal-to-noise ratio. The typical astrometric accuracy per coordinate per visit will range from 11 mas ($m_r$=21) to 74 mas ($m_r$=24).

Although it will still take one or two decades, the combination of many thousands of accurate positions of TNOs and stellar positions from Gaia will certainly move us from a context where observational campaigns for occultations are launched whenever an accurate prediction is obtained to another scenario where we will probably have to select events of interest and/or set up observational networks also allowing a data-driven approach in the study of these distant objects. Whatever the case, the participation of large communities of observers or robotic, dedicated telescopes will be essential.

### 3.3. Expectations of accuracies and number of predictions

The number of predictions varies as a function of the angular diameter of the TNO and mainly of the stellar field that backgrounds the apparent sky path of the TNO. This said, the number of occultation events per TNO is typically 10 per year for a $m_V$=22 target and for stars as faint as $m_V$ ~20, but this depends on the position of the occulting body.

From Fig. 1, it looks safe (and conservative) to say that, after the ten year LSST survey, orbits can be accurate to 50 mas or better (at least until few months after the latest observation of a given ephemeris) for at least several thousand TNOs. For an average of ~10 stellar occultations per year for stars as faint as 20mag, there would be several tens of thousands of predictions. The bottleneck would be the needed updates from 50 mas to 10 mas accuracy. Setting a limit of $m_V$=18 for the stars, the number of reasonably accurate predictions drops to a more tractable value of around a few thousand.

Within a decade or two from now, a first consequence from the big surveys and Gaia could therefore be a significant increase in the number of successful observed occultations, provided that the observational efforts are organized in an effective way, in a similar fashion as asteroidal occultations are being currently organized and observed within amateur networks, but some means of giving priority to specific events should be devised. Scientifically-driven priorities should be stablished in some way.

## 4. The future of the observations

In this section we deal with observing possibilities and interesting opportunities in the future.

### 4.1. Telescope networks

The best performance in the TNO occultation field is achieved with telescope networks that are specifically designed for TNO occultations. If they can be arranged in a large north-south stretch of a good-weather area of the world, the success rate can be good. This is the reason that explains the successes in South America, which represent almost 50% of the published occultations thus far (see tables 1 and 2).

The RECON Research and Education Collaborative Occultation Network (Buie and Keller 2016), is a good example of a specific network created with the main goal of detecting occultations by TNOs. The greatest success of RECON thus far has been the detection of the occultation by 2007UK126, within the context of a broad international collaboration (Benedetti-Rossi et al. 2016). The main current limitation of RECON is the fact that the telescopes are somewhat small and the magnitude limit for the setups is somewhat below 17, which is brighter than the stars in most of our predictions and observations.

Future robotic networks could afford to systematically observe predictions with low probability and given the expected overall future decrease of orbital uncertainties to 50 mas, the probabilities would not be too bad to achieve good results by mere serendipity. On the other hand, contributions have been made by large networks of amateur asteroid occultation observations exist in USA, Europe, Japan, Australia and New Zealand. However, they are devoted to asteroid occultations, most of them using video cameras and telescopes smaller than 30 cm, which limits their observations to a few events involving relatively bright stars ($m_V$<14). New technologies and inexpensive, fast sensitive cameras, with nearly no deadtimes, and capable of integrating, are appearing in the market. Also, larger telescopes may be available to the amateurs in the future.

All this, in combination with portable, deployable networks, can give rise to an increase in the detection rate of occultations. The extremely remarkable success of the occultation by 2014 MU69, the target of the New Horizons space mission, on July 17$^{th}$, 2017 is an example of the state-of-the art in the prediction and observation of stellar occultations by TNOs with the use of deployable networks (Zangari et al. 2018).

**4.2. Large telescopes on Earth**

The two largest optical telescopes planned to be built on Earth are the Extremely Large Telescope (ELT) and the Thirty Meter Telescope (TMT). They may allow recording occultations of extremely faint stars, which implies that many stellar occultations might be detectable per year, but currently there are no stellar catalogs reaching the needed depth and the detections would be single-chord. If occultations by brighter stars near their sites, extremely high signal to noise may be obtained, which is an aspect that can improve the scientific output especially for searches for rings or an atmosphere (see section 5)

**4.3. James Webb Space Telescope**

The James Webb Space Telescope has already specific programs to observe stellar occultations in its GTO programs. The feasibility of observing occultations with HWST has been analyzed in (Santos-Sanz et al. 2016). Although the detections will be single-chord, there will potentially be good science emerging from the infrared capabilities, particularly in atmospheric science.

**4.4. Stratospheric balloon observatories**

Small to medium light-weight telescopes on board stratospheric balloons are being studied as potential future facilities for astronomical use (Maier et al. 2018). If long-lasting flights of around a month or more, launched from several sites on Earth, are possible, occultation events may be monitored even in day light, and the improved seeing at the stratosphere would allow reaching fainter stars than usual. Maybe the highest potential of these instruments is related to the possibility of being deployed within the shadow paths of predicted occultations, as this would be a guarantee against clouds or bad weather ruining the observations, for high scientific value occultations.

**4.5. SOFIA**

The SOFIA airborne observatory has also similar advantages to those mentioned in the previous paragraph. The main limitations are star brightness that can be accessible (because of high turbulence induced by the plane door) and mainly the deployment options and range, as well as its cost. For particularly interesting events, SOFIA can play an important role.

**4.6. Cherenkov Telescopes**

Large telescopes designed to detect the Cherenkov radiation from the avalanche of particles as a result of gamma rays hitting the atmosphere are the so-called Cherenkov Telescopes. They usually have very large collecting areas and cover large fields of views but have poor spatial resolution and cannot observe faint stars. However, they are capable of achieving extremely high time resolution measurements of bright targets. The use of this type of facilities for the study of serendipitous occultations has been assessed in (Lacki 2014) but their use in predicted occultations has not been analyzed in detail yet. Currently, the central detector of telescopes such as the MAGIC telescope at La Palma can observe stars as faint as 13 mag with extremely high time resolution (J. Cortina, personal communication). In the future, this may be improved.

**4.7. Radiotelescopes and radio occultations**

Asteroidal occultations in the radio domain have recently been obtained for the first time

(Lehtinen et al. 2016, Harju et al. 2018). Because the profiles are dominated by diffraction effects at radio wavelengths, the size of the asteroid can be retrieved from a single site. Therefore, the technique may prove useful for TNOs as well. Unfortunately, the number of radio sources in the radio catalogs are still very low compared to the number of stars, but this can change considerably after the Square Kilometer Array (SKA) or its precursors enter operation. Also, the number of powerful enough radiotelescopes on Earth is not comparable to that of optical telescopes, but in the long term the technique may be fruitful for TNOs.

## 5. Aspects needed to improve the scientific output

Even though we have shown the great interest of stellar occultations by TNOs, often the scientific results from the detected events are not the best possible. In this section we deal with aspects that can help to get more and better scientific results.

### 5.1 Efforts for Multiple-chord vs single-chord occultations

Single-chord occultations do not allow us to determine the size of a TNO, but just a lower limit to it (assuming spherical shape). Note that such occultations can also be useful to detect or put stringent limits of an atmosphere. Two-chord occultations can also provide limited results, because we know that, in general, the projected shape of a TNO should be elliptical, not circular. The most likely shape of a TNO is an oblate spheroid, followed, with less probability by triaxial ellipsoids (Duffard et al. 2009) all of which give elliptical projections. A circle can be uniquely determined with 4 points (two chords extremities), but not an ellipse. For a unique elliptical solution at least three chords are needed. Therefore, efforts to achieve multichord events are important. Typically, this requires ~15 participating stations as discussed in previous sections.

### 5.2 Taking advantage of single-chord and two-chord occultations

Single-chord occultations at least provide an accurate position of the TNO to a level below the angular size of the TNO, which helps to improve the orbit of the TNO. But single-chord occultations combined with accurate astrometry of the TNO at the moment of the occultation to pinpoint the centroid position with respect to the chord, can result in at least estimates of the size, as was done in Alvarez-Candal et al. (2014). In order to apply this technique, the telescope used for the occultation observations must be powerful enough to record the occulting target with high signal to noise ratio and the images must have several stars in the field to allow for relative astrometry to be performed. Since the occulting body is usually several magnitudes fainter than the occulted star this requires telescopes of the 2m class and above in practice. Also, regular astrometry, if it can be accurate to the few mas level, can also be used to place the centroid relative to the observed chord so that a size can be estimated assuming a circular projection or an elliptical projection, in a similar way as mentioned above.

### 5.3 Deriving accurate geometric albedos

The geometric albedo of a TNO can be derived once a good projected area has been obtained from an occultation, using this simple equation:

$$p_v = \frac{10^{0.4(H_{sun}-H)}}{(A/\pi)}$$

Where H is the absolute magnitude of the TNO in a specific photometric band, $H_{Sun}$ is the magnitude of the Sun in the same band and A is the projected area in square astronomical units. Since the absolute magnitude H enters in the computation of geometric albedo and H is often not well known from data in the scientific literature, efforts to derive the correct H value are important. Also, it is necessary to discount the contributions of satellites and rings to H. If the satellites are large, this can make a difference of up to several percent. Rings or dust around the body can also make a contribution in reflected light, which must be discounted from the H magnitude. In the case of Haumea this contribution appears to be very small, of the order of 1% only, but on Chariklo and Chiron, their contribution is significant (Duffard et al. 2014, Ortiz et al. 2015).

The tiny contribution of the recently discovered small satellite of Makemake (Parker et al. 2016) to its absolute magnitude does not have an important effect on the albedo, but if there is a yet-undiscovered close satellite to Makemake, the albedo of Makemake could be lower than what has been derived thus far. The two-terrain thermal model (Lim et al. 2010) needing 10% of total Makemake's area in dark terrain might be an indication for such a large satellite.

Also, rotational variability in the value of H must be accounted for because the H value used must be the one at the moment of the occultation. This is especially important for objects with large amplitude of variability. The phase angle law is needed and it is often lacking. Observations of the TNO at different solar phase angles to derive H are usually important. Also, H is a function of epoch, given that the aspect angle changes somewhat in decades, especially for centaurs. This must be modeled with a shape model if data separated by several years is used. For Haumea, all this was taken into account (Ortiz et al. 2017), but this has not often been possible for all the occultations.

A reassessment of all the published geometric albedos of TNOs from stellar occultations is made in table 2, taking all the above effects into consideration. An example of reassessment for 2007 UK126, with the inclusion of a new elliptical fit (combining the data from two different publications) is shown in Fig. 5.

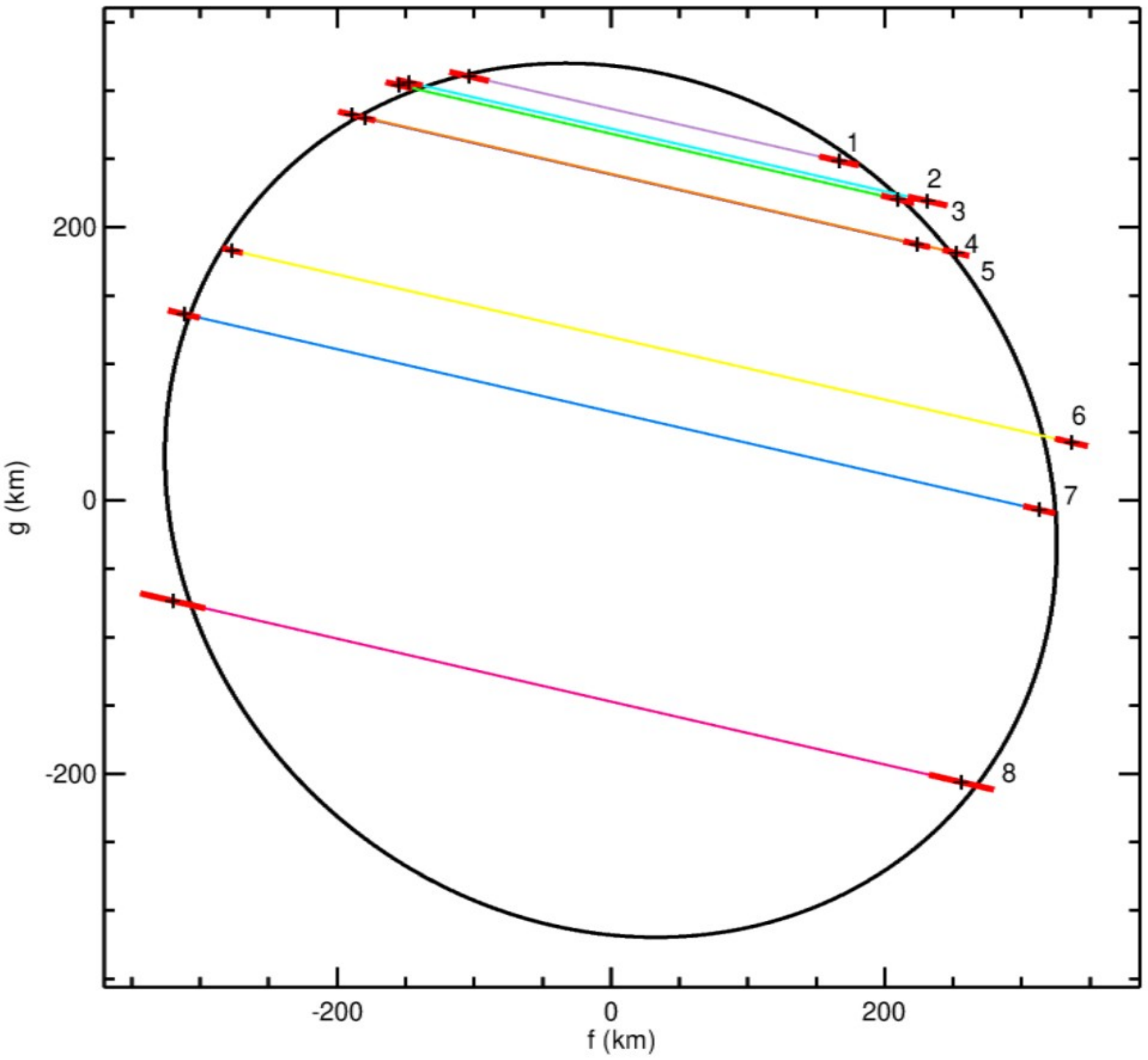

Figure. 5. Chords and new elliptical fit for 2007 UK126, combining the seven chords in Benedetti-Rossi et al. (2016) (marked here as 1.-Reno 2.- Carson City S 3.- Carson City B 4.- Yerington 5.- Gardnerville Nevada 6.- Tonopah 8.- Urbana Illinois) with a non-redundant chord of the 3-chord

event reported in Schindler et al. (2017) for the same occultation. This chord from Sierra Remote Observatories is labeled with 7 here. The red segments indicate the uncertainties in the chords. North is up, East to the left. The new fit (a=678.0 ± 11 km, b= 611.4 ± 18 km), with the new information on absolute magnitude (and taking into account a contribution of the known satellite of 0.036), gives rise to a new geometric albedo of 0.142.

**5.4 Deriving three-dimensional shapes**

For bodies without rotational symmetry, occultations take just a snapshot of the complete shape, at a particular rotational phase. More information from other techniques is needed in order to derive the full three-dimensional shape, or several occultations are needed together with a model shape to determine the 3D shape of the body.

Lightcurve observations of the TNO a few days around the occultation date are often needed so that the extrapolation of the rotation period does not give a large uncertainty in the rotational phase at which the occultation took place. Besides, this kind of observation is also needed to use the proper H value at the time of the occultation for a good geometric albedo estimate, as mentioned above.

**5.5 Timing accuracy**

Unfortunately, telescope equipment that is not specific for occultations can produce inaccurate absolute timings for a variety of reasons. Although efforts are made to avoid this problem, it is sometimes present in the data sets. Then, shifts in time must be applied to the chords of one or several sites. However, this must be done with a suitable methodology. A reasonable method was outlined in Braga-Ribas et al. (2013) to correct for small timing problems, by looking for the best time shifts that aligned the centers of the chords in a straight (but tilted) line.

The case of the stellar occultation by 2002TX300 is a clear example in which a large timing problem of almost 32s was present (Elliot et al. 2010). And unfortunately, only two chords were observed. By shifting the problematic chord as in Elliot et al. (2010) to make their centers coincide the authors were assuming that 2002TX300 would be spherical, whereas we know that this is not the case and the most likely shape of a TNO is an oblate spheroid (Duffard et al. 2009). As a result, the albedo determined by Elliot et al. was too large. An example of an elliptical fit, compatible with the observations and an axial ratio 1.28 is shown in Fig. 6., but implies a larger effective diameter (320 km) than that in Elliot et al. If 2002TX300 has a hydrostatic equilibrium shape, for the rotation rate of ~8h, the minimum axial ratio a/c would be 1.28 for a density around 1000 kg/m$^3$. For a more plausible density of 800 kg/m$^3$ the axial ratio would be 1.39. Alternatively, the 32s difference might be real and could be reflecting a binary body with a configuration as sketched in Fig 6. In this case, a lower limit to the effective diameter would be also ~320 km. This implies a geometric albedo of 0.76 if we use the $H_V$ value of 2002TX300 from Rabinowitz et al. (2008) or 0.68 if we use the same Hv used in Elliot et al. With the latest Hv=3.574 determination by Alvarez-Candal et al. (2016), the resulting geometric albedo would be 0.65. Either in the elliptical case or in the binary scenario this means that 2002TX300 would have an albedo more in accord with the recently determined value for Haumea (0.51), which makes sense given that 2002TX300 is in the family of bodies orbitally related to Haumea.

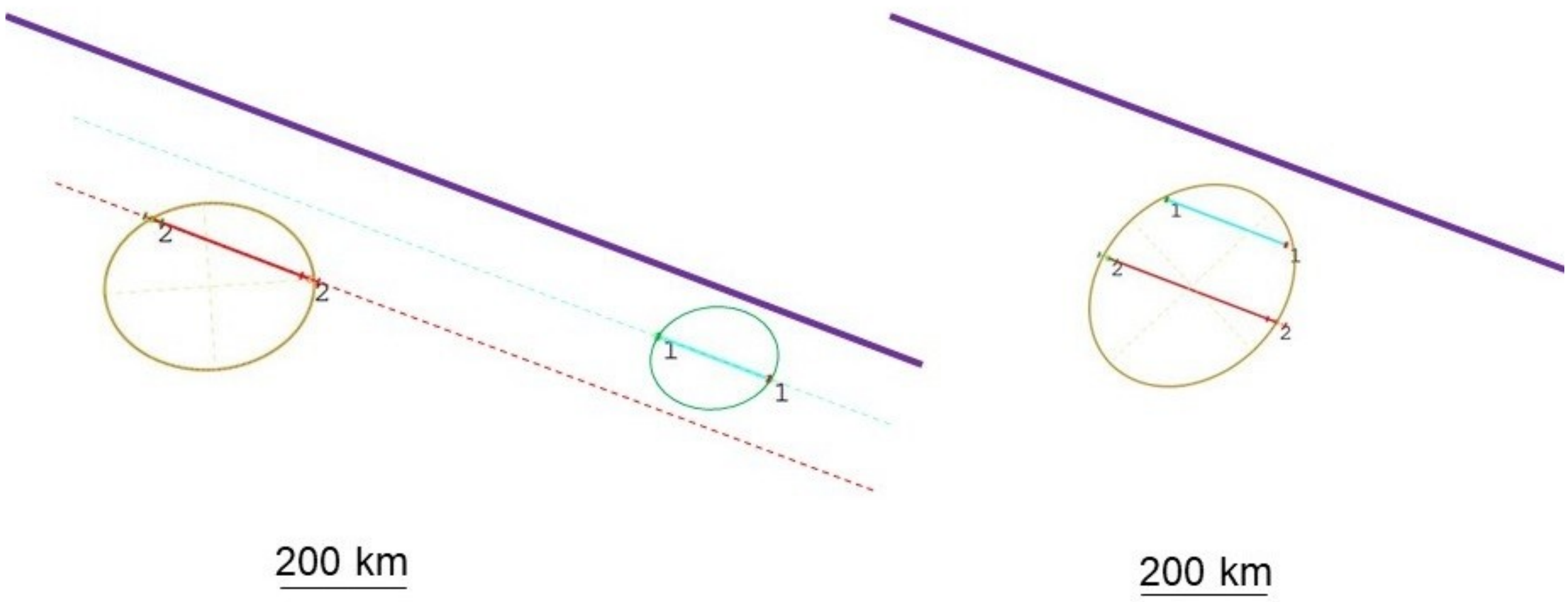

Figure 6. Sketch of two possibilities for fits to the chords of the 2002TX300 event of October 9, 2009. The continuous red and blue lines correspond to the chords in Mauna Kea and Faulkes telescope chords respectively (labeled 2 and 1 respectively). The purple line marks the negative chord at Leeward Community College. The dashed lines on the left panel indicate that just single events would be detected in sites 1 and 2 for the two proposed nuclei. The minimum size for the body is ~320 km. North is up, East to the left.

Also, it is worth mentioning that topographic features can be determined with good accuracy provided that the timings are accurate. Otherwise, some of the deficiencies in the fits translate into residuals that can be wrongly interpreted as topographic features.

**5.6 Detection of satellites and rings**

In order to detect rings, given that they could be as narrow as 1 km, or even less, good temporal sub-second resolutions are needed, with good enough signal to noise ratios. At the typical occultation speed of 20 km $s^{-1}$, a 0.1s integration corresponds to 2 km and depending on the opacity of the ring, the brightness drop may reach less than 50%. Hence, in order to detect rings as thin as 1 km, time resolutions of <0.1s and signal to noise ratios around 10 are needed to detect the events at 5 sigma. This requires fast readout cameras, ideally with no deadtimes and very low readout noise. EMCCD cameras fall in this category but they are usually expensive for large telescope networks. Currently, cameras based on CMOS sensors with extremely low noise levels and fast readout are becoming available at low cost, which will have a positive impact. However, the telescope diameters needed are typically beyond those in the amateur level in order to reach the needed signal to noise. But for bright ~15-mag stars this can be possible with amateur equipment. Detection of satellites around the main occulting bodies is possible too, although of low probability. It may be an important scientific output, but in practice very dense coverage in terms of chords would be needed, which is still very difficult nowadays. In the future, with large and dense networks, this may become possible. This possibility is especially interesting for satellites that are too small and too close to the body to be detectable through imaging even with the JWST.

**5.7. Atmospheres**

Global atmospheres are easy to identify through occultations, as it is the case for Pluto (a topic not covered here). However, local atmospheres are not, because it is not likely that one of the chords will sample, by serendipity, the specific part where the atmospheric dome could be present. Detection is made even harder for an atmospheric dome, since the maximum pressure would be close to the sub-solar point (depending on the thermal inertia, rotation rate, and pole position), and the pressure can be much less at the dawn and dusk limbs. This is quantified in e.g., Young et al. 2017, Ground-based stellar occultations of TNOs essentially only probe the dawn and dusk limbs.

Dense, multichord observations are needed for this. Alternatively, or complementary, central flashes from local atmospheres can potentially be detected provided that one or several of the chords sample a region near the center of the body. From ray-tracing simulations of local atmospheres on non-spherical bodies, one can conclude that some rays can reach the observers near the centrality, and if the object departs considerably from spherical, the regions of the body where anomalous flux can be measured can be larger than just a central spot (Ortiz et al. 2012 supplementary material). Therefore, achieving very high signal to noise ratio at the bottom of the occultation profiles may allow detecting anomalous flux that could arise from such atmospheres. Features of that type were tentatively seen in the Makemake stellar occultation in 2011 although a cosmic ray hit cannot be discarded as a possible source for the main spike (Ortiz et al. 2012). Quaoar may have also exhibited something similar near the centrality in its 2011 occultation, but the signal to noise ratio was not high enough (Braga-Ribas et al. 2013). The largest TNOs are the best candidates to look for such features, if they exist. Spectrally resolved occultations with high cadence spectrographs would also allow detecting absorptions at specific wavelengths, in order to determine gas composition and dust properties.

## Acknowledgements

JLO acknowledges support from grants AYA2017-89637-R and FEDER funds. JLO and PSS acknowledge financial support from the State Agency for Research of the Spanish MCIU through the "Center of Excellence Severo Ochoa" award for the Instituto de Astrofísica de Andalucía (SEV-2017-0709). B.S. acknowledges support from the French grants ‘Beyond Neptune’ ANR-08-BLAN-0177 and ‘Beyond Neptune II’ ANR-11-IS56-0002. Part of the research leading to these results has received funding from the European Research Council under the European Community’s H2020 (2014-2020/ERC grant agreement no. 669416 ‘Lucky Star’). P.S.S. acknowledges financial support by the European Union’s Horizon 2020 Research and Innovation Programme, under Grant Agreement No. 687378. J.I.B.C. acknowledges CNPq grant 308150/2016-3. F.B.R. acknowledges CNPq grant 309578/2017-5. We acknowledge the use of Occult software by D. Herald.

*Note added in proof. Since the writing of this paper, 8 more occultations have been detected in the following dates for the indicated TNOs: 20 Oct 2018, 2015 TG387, 28 Nov 2018, Chiron, 24 Dec 2018 2005 RM43, 30 Dec 2018 2002 WC19, 11 Jan 2019 Bienor, 3 Feb 2019 2005 RM43, 25 Feb 2019, 2015 EZ51, and 18 Mar 2019 Huya.*